\documentclass[twocolumn,showpacs,preprintnumbers,amsmath,amssymb]{revtex4-1}
\usepackage{graphicx}
\usepackage{dcolumn}
\usepackage{bm}

\begin{document}

\title{Intrinsic anomalous Hall effect in nickel: An GGA+U study}
\author{Huei-Ru Fuh and Guang-Yu Guo}\email{gyguo@phys.ntu.edu.tw} 

\affiliation{Department of Physics and Center for Theoretical Sciences, National Taiwan University, Taipei 10617, Taiwan}
\affiliation{Graduate Institute of Applied Physics, National Chengchi Unversity, Taipei 11605, Taiwan}

\date{\today}

\begin{abstract}
The electronic structure and intrinsic anomalous Hall conductivity of nickel have been 
calculated based on the generalized gradient 
approximation (GGA) plus on-site Coulomb interaction (GGA+U) scheme.
The highly accurate all-electron full-potential linearized augmented plane wave method 
is used. It is found that the intrinsic anomalous Hall conductivity ($\sigma_{xy}^H$) obtained 
from the GGA+U calculations with $U = 1.9$ eV and $J=1.2$ eV, is in nearly perfect agreement with 
that measured recently at low temperatures while, in contrast, the $\sigma_{xy}^H$ from the
GGA calculations is about 100 \% larger than
the measured one. This indicates that, as for the other spin-orbit interaction (SOI)-induced 
phenomena in 3$d$ itinerant magnets such as the orbital magnetic magnetization
and magnetocrystalline anisotropy, the on-site electron-electron correlation, though
moderate only, should be taken into account properly in order to get the correct anomalous
Hall conductivity.  The intrinsic $\sigma_{xy}^H$ and the number of valence electrons
($N_e$) have also been calculated as a function of the Fermi energy ($E_F$).
A sign change is predicted at $E_F = -0.38$ eV
($N_e = 9.57$), and this explain qualitatively why the theoretical and experimental
$\sigma_{xy}^H$ values for Fe and Co are positive. It is also predicted
that fcc Ni$_{(1-x)}$Co(Fe,Cu)$_x$ alloys with $x$ being small, would also
have the negative $\sigma_{xy}^H$ with the magnitude being in the range 
of $500\sim 1400$ $\Omega^{-1}$cm$^{-1}$.
The most pronounced effect of including the on-site Coulomb interaction is
that all the $d$-dominant bands are lowered in energy relative to the $E_F$
by about 0.3 eV, and consequently, the small minority spin X$_2$ hole pocket disappears. 
The presence of the small X$_2$ hole pocket in the GGA calculations is attributed 
to be responsible for the large discrepancy in the $\sigma_{xy}^H$ between theory and
experiment. 
\end{abstract}


\maketitle

\section{Introduction}

Anomalous Hall effect (AHE) refers to the transverse charge current generation in solids 
in a ferromagnetic phase by the electric field and has received intensive renewed 
interest in recent years mainly because of its close connection with spin transport
phenomena\cite{Nag10}. 
There are several competing mechanisms proposed for the AHE. Extrinsic mechanisms of skew scattering\cite{Smi55}
and side jump\cite{Ber70} refer to the modified impurity scattering caused the spin-orbit interaction (SOI).
Another mechanism arises from the transverse velocity of the Bloch electrons induced by the SOI, 
discovered by Karplus and Luttinger\cite{Kar54}, and thus is of intrinsic nature.
This intrinsic AHE has recently been reinterpreted in terms of the Berry curvature of the occupied
Bloch states.\cite{Cha96,Jun02,Ono02,Xia10} Furthermore, recent quantitative first-principles studies based on 
the Berry phase formalism showed that the intrinsic AHE is important in various materials.\cite{Yao04}
In particular, in itinerant ferromagnets such as Fe and Co, the intrinsic anomalous Hall conductivity
given by first-principles calculations\cite{Yao04,Wan07} has been found to agree with the experimental 
anomalous Hall conductivity \cite{Dhe67,Miy01,Tia09} within 30 \%, thereby demonstrating the dominance 
of the intrinsic mechanism. 
Nonetheless, the physical origin of the AHE in nickel is still not fully understood. Recent first-principles
density functional calculations with the generalized gradient approximation (GGA) predicted
a large intrinsic anomalous Hall conductivity of -2203 $\Omega^{-1}$cm$^{-1}$ in Ni\cite{Wan07},
which is more than three times larger than the corresponding experimental value of 
-646 $\Omega^{-1}$cm$^{-1}$ \cite{Lav61} at room temperature. In the latest experiment\cite{Ye11},
the intrinsic Hall conductivity was found to be -1100 $\Omega^{-1}$cm$^{-1}$ at low temperatures. 
Though this value is significantly larger than that of the much earlier experiment\cite{Lav61},
it is still only half of the calculated intrinsic Hall conductivity\cite{Wan07}.  

First-principles GGA calculations have been rather successful in describing many physical 
properties such as crystal structure, elastic constant and spin magnetic moment, 
of itinerant ferromagnets Fe, Co and Ni (see, e.g., Ref. \onlinecite{Guo00} and references therein). 
However, GGA calculations fail in describing some relativistic SOI-induced phenomena
in these itinerant magnets.  For example, the theoretical values of orbital magnetic 
moment account 
for only about 50 \% of the measured ones in Fe and Co\cite{Guo97} and the calculated 
magnetocrystalline anisotropy energy
of Ni is even wrong in sign\cite{Guo91}. This failure of the GGA is generally attributed to
its incorrect treatment of the moderate 3$d$ electron-electron correlation in these systems.
Several theoretical methods that 
go beyond the density functional theory (DFT) with the local density approximation
(LDA) or GGA, such as the orbital-polarization correction\cite{Bro85}
and LDA/GGA plus on-site Coulomb interaction U (LDA/GGA+U)\cite{Ani91,Ani93,Czy94} schemes,
have been developed for better description of the SOI-induced phenomena in magnetic solids. 
Indeed, the orbital-polarization correction has been found to bring the calculated orbital
moments in many itinerant magnets such as Fe and Co in good agreement with experiments\cite{Guo97}.
Furthermore, it has been demonstrated that the correct easy axes and the magnitudes of 
the magnetocrystalline anisotropy energy of Fe and Ni can be obtained within the
LDA+U scheme.\cite{Yan01}

Therefore, in this work, we perform GGA+U calculations for nickel to better understand the mechanism of 
its anomalous Hall effect.  We use highly accurate all-electron
full-potential linearized augmented plane wave (FLAPW) method\cite{And75}.
We find that including on-site electron-electron correlation in nickel has significant
effect on its anomalous Hall effect as well as its electronic structure near the
Fermi level. In particular, the calculated anomalous Hall conductivity reduces
from -2200 $\Omega^{-1}$cm$^{-1}$ (GGA) to -1066 $\Omega^{-1}$cm$^{-1}$,
and the latter value is in very good agreement with the measured low temperature 
intrinsic Hall conductivity of -1100 $\Omega^{-1}$cm$^{-1}$ from latest experiments.\cite{Ye11}

This paper is organized as follows. In next section,
we briefly describe how the intrinsic anomalous Hall conductivity is calculated within
the linear-response Kubo formalism as well as the numerical method and computational details
used in the present work. In Sec. III, we first present the calculated anomalous Hall
conductivity, magnetic moments and also relativistic band structure. We then compare our
results with available experiments and also previous calculations. Finally, we make some predictions
about the intrinsic anomalous Hall conductivity for fcc Ni$_{(1-x)}$Co(Fe,Cu)$_x$ alloys with small Co(Fe,Cu)
concentration $x$, within the rigid band approximation.  
In Sec. IV, we summarize the main conclusions drawn from the present work.

\section{THEORY AND COMPUTATIONAL DETAILS}


The intrinsic anomalous Hall conductivity of a solid can be evaluated by using the 
Kubo formula~\cite{Mar00}. The intrinsic Hall effect comes from the 
static limit  ($\omega$=0) of the off-diagonal element of the optical conductivity~\cite{Mar00,Guo05}.
Following the procedure for the calculation of the intrinsic spin Hall conductivity\cite{Guo05}, 
we first calculate the imaginary part
of the off-diagonal element of the optical conductivity 

\begin{equation}
\begin{split}
\sigma_{xy}^{(2)}(\omega)=&\frac{\pi e^2}{\omega V_{c}}\sum_{\bf{k}}\sum_{n \neq n^{'}}(f_{{\bf k}n}-f_{{\bf k}n^{'}})Im[<{\bf k}n|v_{x}|{\bf k}n^{'}>\\
&\times <{\bf k}n|v_{y}|{\bf k}n^{'}>]\delta(\hbar\omega-\epsilon_{n^{'}n})
\end{split}
\end{equation}
where  $V_c$ is the unit cell volume, $\hbar\omega$ is the photon
energy, $|{\bf k}n>$ is the $n$th Bloch state with crystal
momentum ${\bf k}$, $v_{x(y)}$ is the velocity operator, 
and $\epsilon_{n^{'}n} = \epsilon_{{\bf k}n^{'}} - \epsilon_{{\bf k}n}$.
We then obtain the real part from the imaginary part by a Kramers-Kronig transformation
\begin{equation}
\sigma_{xy}^{(1)}(\omega )=-\frac{2}{\pi}\bf{P}\int_{0}^{\infty}d\omega^{'}\frac{\omega^{'}\sigma _{xy}^{(2)}{(\omega^{'})}}{\omega ^{'2}-\omega ^{2}}
\end{equation}
where $\bf{P}$ denotes the principle value of the integral. The intrinsic anomalous 
Hall conductivity $\sigma_{xy}^H$ is the static limit of the off-diagonal element of
the optical conductivity $\sigma_{xy}^{(1)}$($\omega$=0). We notice that the anomalous Hall
conductivity of bcc Fe calculated in 
this way\cite{Guo95,Yao04} is in quantitative agreement with that calculated directly by accounting 
for the Berry phase correction to the group velocity~\cite{Yao04}.

Since all the intrinsic Hall effects are caused by the SOI, first-principles calculations
must be based on a relativistic band theory. Here the relativistic band structure of fcc Ni
is calculated using the highly accurate FLAPW method, 
as implemented in the WIEN2K code\cite{Bla02}.
The self-consistent electronic structure calculations are based on the DFT 
with the GGA for the exchange correlation potential~\cite{Per96}. 
To further take $d$-electron correlation into account, we include on-site Coulomb interaction $U$
in the GGA+U approach\cite{Ani91}. The so-called around-mean-field (AMF) scheme for double counting 
correction\cite{Czy94} is adopted here. $U = 1.9$ eV and $J = 1.2$ eV, 
which were found to give the correct sign and magnitude of magnetocrystalline anisotropy 
energy for fcc Ni\cite{Yan01}, are used. Furthermore, the GGA+U calculations using the
double counting correction scheme designed for approximate self-interaction correction 
(SIC) for strongly correlated systems such as transition metal oxides\cite{Ani93},
have also been performed. Nevertheless, the theoretical anomalous Hall conductivities
from the GGA+U calculations using both double counting correction schemes,
are almost identical (see Table I below). Since the AMF scheme is believed to be more suitable
for metallic systems\cite{Czy94}, here we will concentrate on the results of the AMF calculations.
Moreover, the GGA+U calculations with a larger $U$ value of 2.5 eV are also performed to see how
the variation of $U$ may affect the calculated anomalous Hall conductivity. 

The experimental lattice constant $a = 3.52$ \AA \cite{Kittel} is used here.
The muffin-tin sphere radius ($R_{mt}$) used is 2.2 a.u.
The wave function, charge density, and potential were expanded 
in terms of the spherical harmonics inside the muffin-tin spheres, and
the cutoff angular momentum ($L_{max}$) used is 10, 6, and 6, respectively. 
The wave function outside the muffin-tin spheres was expanded in terms of the augmented plane
waves, and a large number of augmented plane waves (about 70 APWs per atom, i.e., the maximum
size of the crystal momentum $K_{max} = 8/R_{mt}$) were included in the present calculations.
The improved tetrahedron method is used for the Brillouin-zone integration.\cite{Blo94}
To obtain accurate ground state charge density as well as spin and orbital magnetic moments,
a fine $56\times56\times56$ grid of 185193 $k$-points in the first Brillouin zone was used.

\begin{figure}
\includegraphics[width=80mm]{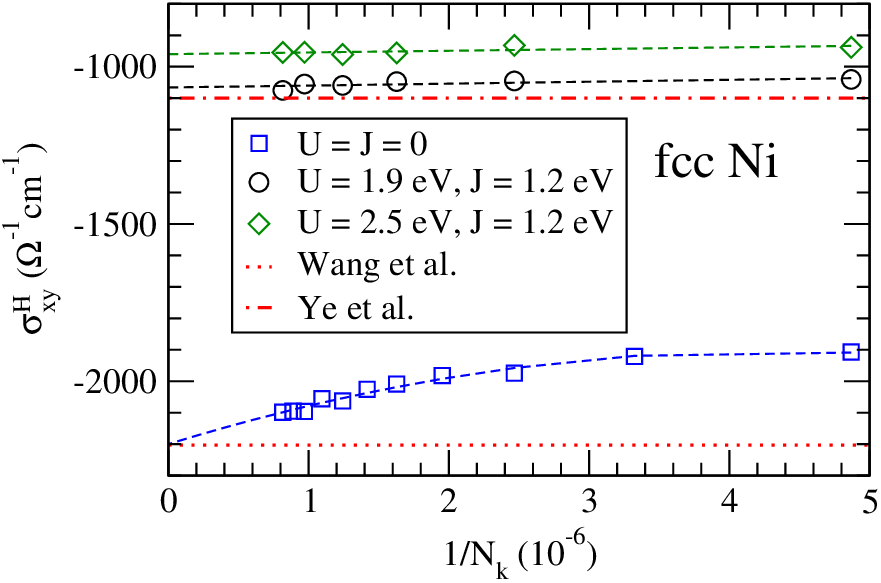}
\caption{(Color online) Calculated anomalous Hall conductivity $\sigma_{xy}^H$ as
a function of the inverse of the number of $k$-points in the Brillouin zone ($N_k$).
The dashed lines are a polynormial fit to the
calculated values to get the extropolated value of $\sigma_{xy}^H$ at $N_k = \infty$.
For comparison, the theoretical value by Wang {\it et al.}\cite{Wan07} and the experimental
value by Ye {\it et al.}\cite{Ye11} are also shown as the horizontal dotted and dot-dashed lines, respectively.}
\label{sigmak}
\end{figure}
 
\section{Results and discussion}
 

Like the calculation of the magnetocrystalline anisotropy energy of bulk magnets\cite{Guo91},
a very fine $k$-point mesh is needed for the anomalous Hall conductivity calculation.\cite{Yao04,Wan07}
Therefore, we perform the $\sigma_{xy}^H$ calculations using several extremely fine $k$-point meshes 
with the finest $k$-point mesh being $106\times106\times106$. The calculated $\sigma_{xy}^H$ 
is plotted as a function of the inverse of the number ($N_k$) of $k$-points in the first Brillouin zone
in Fig. 1. The calculated values of $\sigma_{xy}^H$ are fitted to a polynormal to get the converged
theoretical $\sigma_{xy}^H$ (i.e., the extropolated value of $\sigma_{xy}^H$ at $N_k = \infty$) (see Fig. 1).  
The theoretical $\sigma_{xy}^H$ obtained in this way as well as the calculated spin ($m_s$) and orbital ($m_o$)
magnetic moments are listed in Table I. Also listed in Table I are the available experimental
and previous theoretical $\sigma_{xy}^H$, $m_s$ and $m_o$. 

Table I shows that the theoretical $\sigma_{xy}^H$ from the GGA is -2200 $\Omega^{-1}$cm$^{-1}$,
and is much larger than the experimental value of -646 $\Omega^{-1}$cm$^{-1}$ at room 
temperature\cite{Lav61} and also the experimentally derived intrinsic value 
of -1100 $\Omega^{-1}$cm$^{-1}$ at low temperatures\cite{Ye11}.
Nevertheless, the present result is in very good agreement with the previous GGA 
calculations\cite{Wan07}, as it should be (see Table I and Fig. 1). 
Interestingly, when the on-site Coulomb interaction is taken into account via the
GGA+U scheme, the calculated $\sigma_{xy}^H$ reduces significantly.  
In particular, when $U = 1.9$ eV and $J = 1.2$ eV which were found to give rise to
the correct easy axis and magnitude of the magnetocrystalline anisotropy energy\cite{Yan01},
the theoretical  $\sigma_{xy}^H$ becomes -1066 $\Omega^{-1}$cm$^{-1}$.
This good agreement between the present GGA+U calculation and the low temperature
measurements\cite{Ye11} (Table I and Fig. 1) indicates that the intrinsic AHE dominates in nickel and
that the 3$d$ electron-electron correlation in itinerant magnets such as nickel has an important
effect on the AHE, although being only moderate. 
Further increasing $U$ to 2.5 eV reduces the calculated $\sigma_{xy}^H$ slightly (Table I and Fig. 1).
As expected, the GGA+U calculations increase the theoretical spin and orbital magnetic moments,
and hence enlarge somewhat the discrepancies between the calculations and experiments (Table I). 
Nonetheless, with $U = 1.9$ eV and $J = 1.2$ eV, the theoretical spin and orbital magnetic moments
are still in reasonable agreement with the experiments (Table I). 

\begin{table}
\caption{Calculated anomalous Hall conductivity $\sigma_{xy}^{H}$ ($\Omega^{-1}$cm$^{-1}$) as well as
spin magnetic moment $m_s$ ($\mu_B$/atom) and orbital magnetic moment $m_s$ ($\mu_B$/atom). 
Superscripts of SIC indicate that the values are obtained from the GGA+U calculations 
with the SIC double counting correction scheme (see Sec. II).
The corresponding experimental values as well as previous theoretical $\sigma_{xy}^{H}$ are 
also listed for comparison.}
\begin{ruledtabular}
\begin{tabular}{ccccc}
               &GGA         &GGA+U    &GGA+U          & Expt.  \\ 
               &            & U=1.9eV & U=2.5eV    &      \\  \hline
$\sigma_{xy}^H$&-2200       &-1066         & -960         & -646$^a$ \\
               &-2203$^b$   &-1107$^{SIC}$ & -945$^{SIC}$ & -1100$^c$ \\
$m_s$          & 0.639      &0.661         & 0.675        & 0.57$^d$ \\
               &            &0.685$^{SIC}$ & 0.707$^{SIC}$ &      \\
$m_o$          & 0.051      &0.066         & 0.071         & 0.05$^d$ \\
               &            &0.069$^{SIC}$ & 0.076$^{SIC}$ &      \\
\end{tabular}
\end{ruledtabular}
\noindent $^a$Reference \onlinecite{Lav61}.\\
$^b$Reference \onlinecite{Wan07}.\\
$^c$Reference \onlinecite{Ye11}.\\
$^d$Reference \onlinecite{Lan88}.
\label{moment}
\end{table}


\begin{figure*}
\includegraphics[width=120mm]{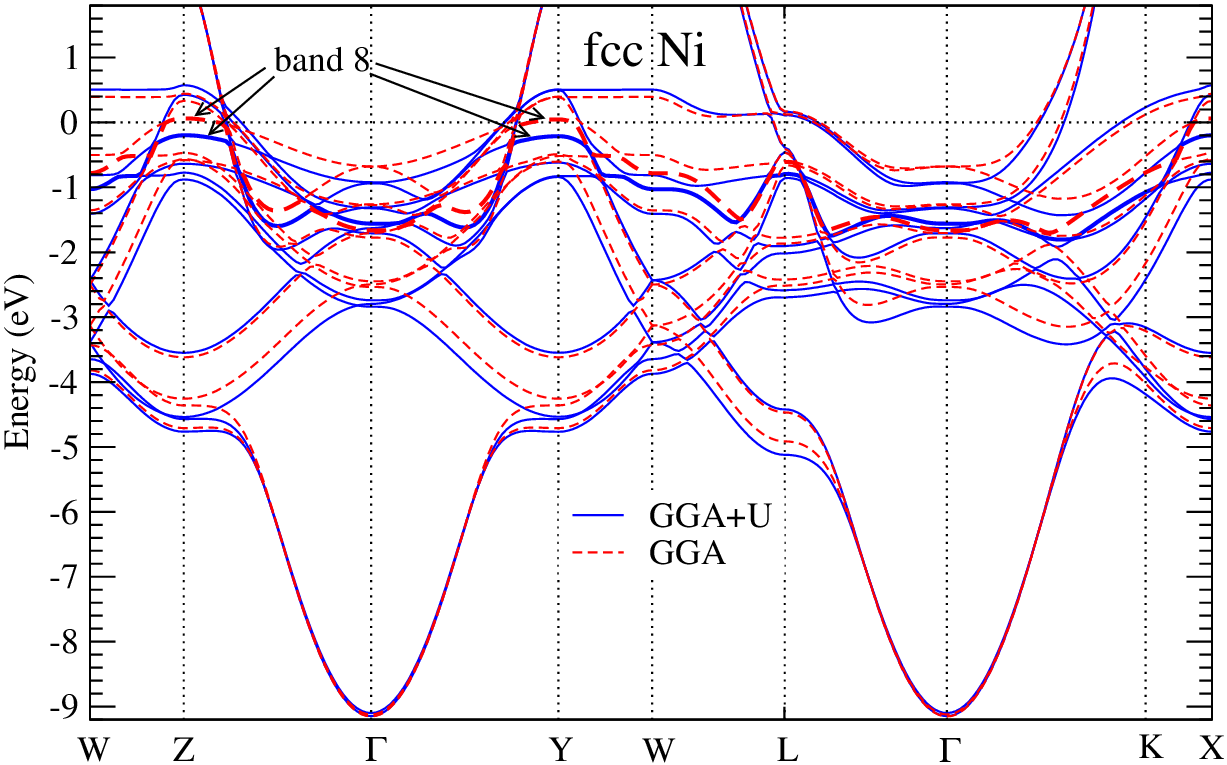}
\caption{(Color online) Relativistic GGA and GGA+U band structures. The GGA+U band structure was
obtained using $U = 1.9$ eV and $J=1.2$ eV.
The Fermi level (dotted horizontal line) is at 0 eV. The thick curves
denote band 8's which are also indicated by the arrowed lines.}
\label{bands}
\end{figure*}

To help understand how the on-site electron-electron correlation affects the electronic 
band structure and AHE in nickel, we plot in Fig. 2 and Fig. 3 the relativistic energy bands along the 
high symmetry lines in the Brillouin zone calculated both without and with on-site 
Coulomb interaction U. 
The relativistic band structure may be regarded as
the result of a superposition of the corresponding scalar-relativsitic spin-up 
and spin-down band structures with many accidental band-crossings (degeneracies)
lifted by the SOI. In nickel, nevertheless, these SOI-induced splittings are generally much 
smaller than the exchange splittings, and thus can be treated as a perturbation.\cite{Wan74} 
Also, including the SOI would lower the symmetry of the system. In the present work,
the magnetization is assumed to be along the [001] direction, and the symmetry of the system
becomes the tetragonal one. The six high symmetry X points which are equivalent
in the nonrelativistic (or scalar-relativistic) case, now form two inequivalent groups,
namely, two equivalent $\pm$ Z points and four equivalent $\pm$ X and $\pm$ Y points.
Therefore, the energy bands near the Z point are slightly different from that near
the X and Y points, as shown in Fig. 2 and Fig. 3. 
The relativistic band structure and Fermi surface of Ni has been reported by
several researchers before by using different band structure calculation methods 
(see, e.g., Refs. \onlinecite{Wan74,Guo91,Yan01,Wan07}).
In particular, Wang and Callaway analysed in detail the energy band characters and Fermi surface
sheets.\cite{Wan74} The present GGA relativistic band structure (the red dashed curves in Fig. 2 and Fig. 3)
is very similar to that reported by Wang and Callaway\cite{Wan74}.
For example, in both cases, there are five bands (bands 8-12) crossing the
Fermi level, and there is a small down-spin-dominant hole pocket (band 8, noted as X$_2$)
centered at the X(Y,Z) symmetry point (see Figs. 2-3, and also Fig. 1 in Ref. \onlinecite{Wan74}).  
However, the small X$_2$ hole pocket was not found in the de Haas-van Alphen experiments.\cite{Tsu67}

Now let us focus on the changes in the relativistic band structure caused by including 
the on-site Coulomb interaction.
A pronounced change is that all the $d$-dominant bands are lowered in energy relative
to the Fermi level, when the on-site Coulomb interaction is taken into account
in the GGA+U scheme (see Figs. 2-3). In other words, the binding energy of the $d$-dominated
valence bands is increased by about 0.3 eV. In particular, the down-spin band 8
is now pushed completely below the Fermi level and the small X$_2$ hole pocket disappears.
The absence of the X$_2$ hole pocket is consistent with the de Haas-van Alphen experiments.\cite{Tsu67}
In Ref. \onlinecite{Yan01}, the absence of the X$_2$ hole pocket caused by the on-site
Coulomb interaction was regarded as the main reason that the LDA+U calculations predicted
the correct easy axis for nickel. 
As will be shown below, this band 8 near the Fermi level calculated without the on-site 
Coulomb interaction, gives rise to a pronounced contribution to the anomalous Hall conductivity, and thus 
the absence of the small X$_2$ hole pocket in the GGA+U band structure is also the
main cause for the significant reduction of the calculated anomalous Hall conductivity.


In Fig. 3, we display the anomalous Hall conductivity ($\sigma_{xy}^H$) and
the number of valence electrons ($N_e$) as a function of the Fermi level ($E_F$),
together with the relativistic band structure, from the GGA+U calculations 
using $U = 1.9$ eV and $J=1.2$ eV. The fine $k$-point mesh of $106\times106\times106$
is used. Clearly, the magnitude of the $\sigma_{xy}^H$ peaks just above the
true Fermi level ($E_F = 0$ eV), with a large value of -1420 $\Omega^{-1}$cm$^{-1}$
at $0.16$ eV ($N_e = \sim 10.3$). The peak may be related to the flat band (band 10) 
along the W-L-$\Gamma$ line near the L point just above the Fermi level (see Fig. 3a).
As the Fermi level is further artificially raised, the size of the $\sigma_{xy}^H$ 
decreases steadily and becomes rather small (within 200 $\Omega^{-1}$cm$^{-1}$)
above 0.75 eV. When the $E_F$ is artificially lowered, the magnitude of the $\sigma_{xy}^H$
initially decreases gradually, and then drops sharply starting at $E_F = -0.35$ eV. 
The resultant shoulder at -0.35 eV can be attributed to the presence
of the top segment of band 8 near the X (Y,Z) point in this energy region (see Fig. 3a). 
The $\sigma_{xy}^H$ then changes its sign at -0.38 eV (or $N_e = \sim9.57$).
As the $E_F$ is further lowered, the  $\sigma_{xy}^H$ increases sharply and then
peaks at -0.74 eV (or $N_e = \sim8.90$) with a large value of 2635 $\Omega^{-1}$cm$^{-1}$.
Beyond this point, the $\sigma_{xy}^H$ decreases and fluctuates but remains positive as the
$E_F$ is further lowered, and it then changes its sign again at -2.28 eV (or $N_e = \sim4.78$).
For the $E_F$ being below -3.80 eV, the magnitude of the $\sigma_{xy}^H$ is small.
Note that the energy bands below -4.0 eV and above 0.5 eV are predominantly of
$4s4p$ character and thus the effects of the SOI and exchange interaction
are small.

\begin{figure*}
\includegraphics[width=120mm]{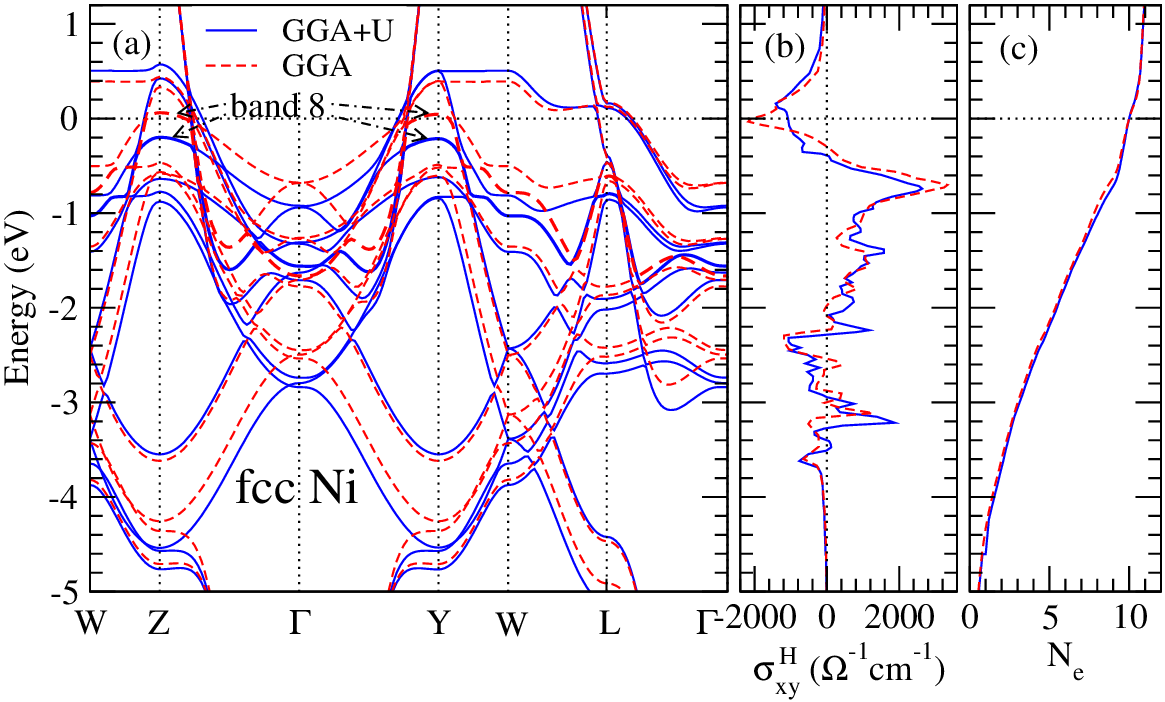}
\caption{(Color online) Relativistic band structure (a), anomalous Hall conductivity 
($\sigma_{xy}^{H}$) (b), and number of valence electrons ($N_e$) (c), calculated with 
(solid blue lines) and without (dashed red lines) on-site Coulomb interaction $U$.
Both $\sigma_{xy}^{H}$ and $N_e$ were calculated using the fine $k$-point mesh 
of $106\times106\times106$. The GGA+U results were obtained using $U = 1.9$ eV and $J=1.2$ eV.
The Fermi level (dotted horizontal line) is at 0 eV.
In (a), the thick curves denote band 8's which are also indicated by the arrowed lines.}
\label{sigmaw}
\end{figure*}

Experimentally, the measured anomalous Hall resistivity $\rho_{xy}$ 
is often analysized in terms of two distinctly different resistivity
($\rho_{xx}$)-dependent terms\cite{Nag10}, i.e., $\rho_{xy} = a\rho_{xx} + b\rho_{xx}^2$.
Since usually $\rho_{xy} \ll \rho_{xx}$, the total anomalous Hall 
conductivity $\sigma_{xy} = -\rho_{xy}/(\rho_{xy}^2+\rho_{xx}^2) \approx a\sigma_{xx} + b$,
where the first linear $\sigma_{xx}$-dependent term ($a\sigma_{xx}$) 
was attributed to the extrinsic skew scattering mechanism ($\sigma_{xy}^{SK}$)\cite{Smi55}. 
The skew scattering contribution has been found to become dominant in dilute impurity metals
at low temperatures.\cite{Nag10} Indeed, recent {\it ab initio} calculations for
the alloy systems Fe$_{1-x}$Pd$_{x}$ and Ni$_{1-x}$Pd$_{x}$ indicated that in the small
Pd concentration $x$ region,  the $\sigma_{xy}^{SK}$
can be several times larger than the $b$, and can also differ in sign.\cite{Low10} 
The second scattering-independent term $b$ was usually further separated 
into the intrinsic contribution $\sigma_{xy}^H$\cite{Kar54} which can be obtained
from band structure calculations\cite{Nag10}, as done here for nickel, 
and also the extrinsic side jump mechanism ($\sigma_{xy}^{SJ}$)\cite{Ber70}.
Therefore, when one compares the calculated $\sigma_{xy}^H$ with the experimental
scattering-independent term $b$, one should be aware of the possible side jump 
contribution $\sigma_{xy}^{SJ}$ even though it has been shown that the $\sigma_{xy}^H$
would dominate in ferromagnet metals such as Fe and Co\cite{Yao04,Wan07}. 
Recent theoretical calculations for the 2D Rashba and 3D Luttinger Hamiltonians 
using a Gaussian disorder model potential suggested that
the AHE in the (III,Mn)V ferromagnetic semiconductors at low temperatures could be
dominated by the $\sigma_{xy}^{SJ}$.\cite{Kov10} However, more recent {\it ab initio} calculations
show that in fcc Fe$_{1-x}$Pd$_{x}$ and Ni$_{1-x}$Pd$_{x}$, the $\sigma_{xy}^{SJ}$ is generally
two-order of magnitude smaller than both the $\sigma_{xy}^{SK}$ and $\sigma_{xy}^H$,
thus being negligible.\cite{Low10} This is consistent with the recent experimental finding of
the negligible $\sigma_{xy}^{SJ}$ in bcc Fe\cite{Tia09} and fcc Ni\cite{Ye11} using
a newly established empirical $\sigma_{xx}$-scaling formula for $\sigma_{xy}$\cite{Tia09}. 

Let us now return to the calculated $\sigma_{xy}^H$ as a function of the $E_F$ described above and
also presented in Fig. 3. We find that it can explain qualitatively within the rigid band approximation
the known AHE experiments on 3$d$ transition metal ferromagnets and
their alloys.\cite{Dhe67,Miy01,Tia09,Lav61} For example, the sign of the $\sigma_{xy}^H$
of both Fe and Co were found to be positive.\cite{Dhe67,Miy01,Tia09} Interestingly,
the calculated $\sigma_{xy}^H$ at $N_e = 8.0$ ($E_F = -1.02$ eV) is 758 $\Omega^{-1}$cm$^{-1}$,
being in very good agreement with the previous {\it ab initio} calculations for Fe\cite{Yao04}.
Of course, this nearly perfect agreement may be accidental because
Fe crystallizes in the bcc structure and also has a much larger spin magnetic
moment, and hence the rigid band model should not work very well here.
We note that the calculated $\sigma_{xy}^H$ is 2548 $\Omega^{-1}$cm$^{-1}$ 
at $N_e = 9.0$ ($E_F = -0.71$ eV), which is much larger than the previous theoretical
$\sigma_{xy}^H$ value for fcc Co\cite{Rom09} and also the experimental $\sigma_{xy}^H$ 
value for polycrystalline hcp Co\cite{Kot05}.  
Nickel forms substitutional alloys in fcc structure with low concentration
of Fe, Co and Cu. The present calculations (Fig. 3) predict that
fcc Ni$_{(1-x)}$Co(Fe,Cu)$_x$ alloys ($x$ being small) would have
the negative intrinsic $\sigma_{xy}^H$ with the magnitude being in 
the range of $500\sim 1400$ $\Omega^{-1}$cm$^{-1}$.
Of course, when comparing this prediction with the AHE experiments
on fcc Ni$_{(1-x)}$Co(Fe,Cu)$_x$ alloys, one should take into accoint
the non-negligible skew scattering contribution $\sigma_{xy}^{SK}$\cite{Low10}.
Indeed, as mentioned earlier, Fig. 3b shows a sign change of $\sigma_{xy}^H$ 
at $N_e \approx 9.570$ ($E_F = -0.38$ eV), whereas the sign change of the $\sigma_{xy}$
in fcc Ni$_{1-x}$Fe$_x$ alloys with $x \approx 0.13$ (i.e., $N_e \approx 9.74$)
was experimentally found\cite{Smi55}. This experimentally found sign change 
at the lower Fe concentration could be attributed to the presence of the positive
$\sigma_{xy}^{SK}$ in fcc Ni$_{1-x}$Fe$_x$ alloys.

As mentioned before, the most pronounced effect of including the on-site Coulomb interaction
is that all the $d$-dominant bands are lowered in energy relative to the $E_F$
by about 0.3 eV. Consequently, the down-spin band 8 is now pushed completely below 
the $E_F$ and the small X$_2$ hole pocket disappears. Now the top of band 8 is
located at -0.20 eV on the X(Y,Z) point (Figs. 2-3). To see clearly the effect of the 
on-site Coulomb interaction on the $\sigma_{xy}^H$, 
we also display in Fig. 3 the anomalous Hall conductivity ($\sigma_{xy}^H$) and
the number of valence electrons ($N_e$) as a function of $E_F$,
together with the relativistic band structure, from the GGA calculations using 
the fine $k$-point mesh of $106\times106\times106$. 
Strikingly, the $\sigma_{xy}^H$ exhibits a pronounced negative peak right at the $E_F$
with the value being about -2178 $\Omega^{-1}$cm$^{-1}$ (see Fig. 3b). The remarkable
difference in the $\sigma_{xy}^H$ at 0.0 eV from the GGA and GGA+U calculations 
can be attributed to the presence of the above mentioned X$_2$ hole pocket in the
GGA calculation (Fig. 3). Another major difference is the absence of the clear 
shoulder at -0.35 eV in the GGA $\sigma_{xy}^H$ spectrum. 
This is because of the absence of the top segment of band 8 near the X (Y,Z) point 
in this energy region in the GGA band structure (see Fig. 3a). In fact, Fig. 3 shows 
that as the $E_F$ is lowered from the true Fermi level (0 eV), the GGA $\sigma_{xy}^H$
decreases in magnitude steeply to zero and changes its sign at -0.30 eV ($N_e = 9.55$).  
Nevertheless, in both cases, the $\sigma_{xy}^H$ changes its sign near $N_e = 9.56$.
The other pronounced difference (by about 700 $\Omega^{-1}$cm$^{-1}$)
in the two calculated $\sigma_{xy}^H$ spectra 
is the height of the positive peak near -0.70 eV. In the rest of the energy region,
the two $\sigma_{xy}^H$ spectra look rather similar.

\section{Conclusions}
 
In summary, we have calculated the electronic structure
and intrinsic anomalous Hall conductivity of nickel with both the GGA and GGA+U schemes.
The highly accurate all-electron FLAPW method is used. We find that the theoretical
anomalous Hall conductivity ($\sigma_{xy}^H$) obtained from the GGA calculations
(Table I) is about 100 \% larger than the intrinsic $\sigma_{xy}^H$ recently measured at low
temperatures\cite{Ye11}. In contrast, the theoretical $\sigma_{xy}^H$ from the
GGA+U calculations with $U = 1.9$ eV and $J=1.2$ eV is in almost perfect agreement with
the measured one (Table I). This indicates that, as for other SOI-induced magnetic
phenomena in 3$d$ itinerant magnets such as the orbital magnetic magnetization 
and magnetocrystalline anisotropy, the on-site electron-electron correlation, though
moderate only, should be taken into account properly in order to get the correct anomalous
Hall conductivity. The most significant effect of including the on-site Coulomb interaction
is that all the $d$-dominant bands are lowered in energy relative to the $E_F$
by about 0.3 eV. Consequently, the down-spin band 8 is now pushed completely below 
the $E_F$ and the small X$_2$ hole pocket disappears. The presence of the
small X$_2$ hole pocket in the LDA and GGA calculations is found to be the
main reason why the large discrepancy in the $\sigma_{xy}^H$ between theory and
experiment exists. The intrinsic anomalous Hall conductivity ($\sigma_{xy}^H$) has also
been calculated as a function of the Fermi level $E_F$ (or the number of valence electrons 
$N_e$) in the GGA+U scheme. A sign change is predicted at $E_F = -0.38$ eV 
(or $N_e = 9.57$), and this explain qualitatively why the theoretical and experimental
$\sigma_{xy}^H$ values for Fe and Co are positive. Finally, the present calculations 
(Fig. 3) indicate that fcc Ni$_{(1-x)}$Co(Fe,Cu)$_x$ alloys ($x$ being small) would also 
have the negative intrinsic $\sigma_{xy}^H$ with the magnitude being in
the range of $500\sim 1400$ $\Omega^{-1}$cm$^{-1}$.

\section*{Acknowledgments}
G.Y.G thanks Xiaofeng Jin for communicating their experimental results\cite{Ye11} prior to
the submission for publication.
The authors acknowledge the supports from the National Science Council and NCTS of Taiwan.
Some calculations were carried out on the computers at the National Center for 
High-Performance Computing of Taiwan.

 

\end{document}